\documentclass[twocolumn, a4paper, twoside, showpacs, amsmath, amsfonts, superscriptaddress, nobibnotes]{revtex4}
 \pdfoutput=1 

\usepackage[utf8]{inputenc}
\usepackage[T1]{fontenc}
\usepackage{graphicx}
\usepackage{stmaryrd} 

\usepackage{pxfonts} 
\usepackage{hyperref}

\usepackage{color}
\usepackage{datetime}

\begin{document}

\newcommand\ket[1]{\left|#1\right\rangle}
\newcommand\bra[1]{\left\langle#1\right|}
\newcommand\ketbra[1]{\ket{#1}\bra{#1}}
\newcommand\braket[2]{\left\langle #1 | #2 \right\rangle}
\newcommand{\eq}[2]{\mathrel{\operatorname*{=}_{#1}^{#2}}}
\newcommand\isbydef{\eq{\text{def}}{}}
\newcommand\abs[1]{\left|#1\right|}

\newcommand\Id{{\mathbb I}}
\newcommand{\Proba}[1]{{\mathcal P(#1)}}
\newcommand\Pdelta{{\mathcal P(\Delta |n,m)}} 
\newcommand{\Hamming}[1]{\left\lVert #1 \right\rVert}
\newcommand{\M}{\mathbb M}

\newcommand{\ie}{\emph{i.e.} }
\newcommand{\eg}{\emph{e.g.} }
\newcommand{\etal}{\emph{et al.}}
\newcommand{\dsp}{DSP }
\newcommand{\irud}{IRUD }
\newcommand{\pns}{PNS }
\newcommand{\wcp}{WCP }
\newcommand{\ud}{UD }

\title{Photon-Number-Splitting-attack resistant Quantum Key Distribution Protocols without sifting}
\author{Fabio Grazioso}
\email{grazioso@iro.umontreal.ca}
\affiliation{Laboratoire de Photonique Quantique et Moléculaire, ENS de Cachan, UMR CNRS 8537, 94235 Cachan cedex, France}
\affiliation{Université de Montréal, Département IRO, CP 6128, Succ. Centre–Ville, Montréal (QC),
H3C 3J7 Canada}
\author{Frédéric Grosshans}
\email{frederic.grosshans@u-psud.fr}
\affiliation{Laboratoire de Photonique Quantique et Moléculaire, ENS de Cachan, UMR CNRS 8537, 94235 Cachan cedex, France}
\affiliation{Laboratoire Aimé Cotton, UPR CNRS 3321, Orsay, France}

\begin{abstract}
  We propose a family of sifting-less quantum-key-distribution protocols which use reverse-reconciliation, and are 
   based on weak coherent pulses (WCPs) polarized along $ m $ different directions. 
  When $ m=4 $, the physical part  of the protocol is identical to most experimental implementations of BB84
  \cite{BB84} and SARG04 \cite{SARG1,SARG2} protocols 
  and they differ only in classical communications and data processing.
  We compute their total keyrate as function of the channel transmission $T$, using general information
  theoretical arguments and we show that they have a higher keyrate than the more standard protocols, both for
  fixed and optimized average photon number of the WCPs.
  When no decoy-state protocols (DSPs) \cite{Hwang-03, LoMaChen-05, Wang-05} are applied, the scaling of the
  keyrate with transmission is improved from $T^2$ for BB84 to  $T^{1+\frac1{m-2}}$. 
  If a DSP is applied, we show how the keyrates scale linearly with $T$, with an improvement of the prefactor
  by 75.96\% for $m=4$.
  High values of $ m $ allow to asymptotically approach the keyrate obtained with ideal single photon pulses. 
   The fact that the keyrates of these sifting-less protocols are higher compared to those of the
   aforementioned more standard protocols show that the latter are not optimal, 
   since they do not extract all the available secret key from the experimental correlations. 
\end{abstract}

\date{\today}


\maketitle

Over the last  three decades, quantum key distribution (QKD)  
has emerged as the main application of quantum information 
\cite{BB84,SBPCDLP09}.
In most experimental realizations,
the legitimate partners --- traditionally named Alice and Bob ---
use the BB84 protocol with weak-coherent-pulses (WCP), \ie
Alice sends polarized coherent states  to Bob,
and Bob measures their polarization to obtain the raw-key.
Alice and Bob then post-select a subset of the measurement to obtain the 
sifted-key from which the cryptographic key is extracted.
If, for each pulse, Eve would send a single photon, 
there would be no way for an eavesdropper
%
%
--- traditionally named Eve --- to learn anything about the sifted key without introducing errors. 
But, with WCP$\,$s, Alice only approximates single-photons, and she sometimes sends
multiphoton pulses, on which Eve can get all the information through 
photons-number splitting (PNS) attack \cite{Luetkenhaus00}.
SARG04 \cite{SARG1, SARG2} showed that, 
with the same physical states  and the same detection of BB84, 
and changing only the post-processing it is possible to enhance the robustness toward PNS attacks. 
 This was obtained by changing the encoding of the raw key bits in the qubits, changing the sifting 
 in such a way that  Eve only gains partial 
information from 2-photon-pulses and needs to wait for the 
rarer 3-photon-pulses to gain the full information. 
However, for the same pulse intensity, SARG04's rate is the half of BB84 at low losses, 
because of its lower sifting rate. 
As shown in \cite{SARG2} SARG04's robustness can be increased by using $m$ polarizations
 instead of 4, at the price of a lower sifting rate $\propto m^{-3}$.
 Similarly, greater values of $m$ give the same increased robustness to our
protocol while essentially keeping the relatively high rate of BB84. 

These examples show that it is possible to improve the performance of a QKD
protocol by changing the classical data post-processing without changing its
physical implementation. The work presented here is mainly motivated by the 
research of the optimal way to extract a secret key from a BB84-like protocol
with WCPs of average photon number $\mu$. 
Although we do not compute an upper-bound of the amount of private information 
one can extract from such an implementation, we propose a family of protocols
without sifting, which have better performances than BB84 and SARG with the same setup.
This sifting step in BB84 and SARG04 can be seen as an arbitrary
choice of encoding, and its absence allows us to describe the protocols more 
directly using information theoretic quantities. 
%
%
These protocols, similar to 
continuous variable protocols \cite{GrosshansGrangier02-Proc,GVAWBCG03}
 seem to perform better than both BB84 and SARG04, 
at least close to the ideal regime studied here, where  errors-rate, finite-size effects 
and key-extraction algorithm imperfections are negligible. 

%
Section \ref{secProtocol} we describe this protocol family;
in particular, the protocol itself is described in subsection \ref{subsecProt}, 
and the keyrate in the ideal case with perfect single photons is computed in subsection \ref{cleanKeyrate}.
In  section \ref{secEve}  we  first introduce some notation ( subsection \ref{appendix-rewriting-states}),
used later to describe  the two kind of errorless attacks 
(\emph{Intercept - Resend with Unambiguous Discrimination} (IRUD) attacks in \ref{subsecIRUD}
 and PNS attack in \ref{subsecPNS}). 
The optimal combination of these attacks is analyzed in \ref{secEavesStrat}, both 
for implementation without (\ref{subsecKnoDSP}), and with decoy states protocols
 (DSP)  (\ref{subsecKDSP}), and finally in \ref{comparisons-section} we compare the sifting-less protocol with BB84.
The conclusion of this analysis is drawn in section  \ref{secConclusion}. 

\section{Description of the protocol family}
\label{secProtocol}

\subsection{Description of the protocols}
\label{subsecProt}
Each protocol of the family is characterized by the number $m\ge3$,
the total number of 
possible polarizations used in the protocol.
To simplify the analysis, we will suppose that
the polarizations are uniformly distributed along the great circle of 
Poincaré's sphere representing the states with linear polarizations. 
Alice randomly choses $x\in\llbracket0,m-1\rrbracket$ and sends 
a phase-randomized weak coherent pulse (WCP) linearly polarized
in the direction $x\theta_m$, with 
\begin{equation}
    \theta_m\isbydef\frac{2\pi}{m}.
\end{equation}
The quantum state sent by Alice will depend on the number of photons $n$ contained
in this pulse.
If $\ket0$ and $\ket1$ are the two states of circularly polarized single photons,
the state of a $n$-photon-pulse polarized along $x\theta_m$ is
\begin{equation}\label{statextheta}
\ket{x,n,m}\isbydef\ket{x\theta_m}^{\otimes n},
\end{equation}
with
\begin{align}
  \ket\theta &
  \isbydef\tfrac1{\sqrt2}\left(\ket0 + e^{i\theta}\ket1\right), 
  &
  \ketbra{\theta}&=\frac12 
    \begin{bmatrix}
      1 &e^{-i\theta} \\
      e^{i\theta}  &1 
    \end{bmatrix}.
    \label{eq:rhok}
\end{align} 

If $m=4$, one has the 4 states used in BB84 \cite{BB84}, 
SARG04 \cite{SARG1,SARG2} as well as LG09 \cite{LeverrierGrangier09}.
We chose to study the generic $m$-state case because it does not change 
the complexity of the analysis.

Bob measures the polarization of the pulses after a propagation into 
an attenuating channel characterized by its transmission $T$.
The public comparison of a small subset of the measurements
allows Alice and Bob to statistically determine the characteristic of the channel, 
namely its yield $Y$ -- the probability for Bob to get a click --
and its qubit error rate (QBER).
In this first analysis, we will suppose this statistical evaluation to be exact, 
neglecting the finite size effects \cite{CaiScarani09}.
We will also limit ourselves to the errorless case, where the QBER is 0,
excepted in the conclusion where the influence of errors is briefly studied.

 In the following we will also denote with $X$ (respectively $Y$)  the classical variables representing the bit values encoded by Alice (respectively measured by Bob). The ambiguity of $Y$ being also the total transmission yield should be clarified by the context, and will be explicitly resolved otherwise.
$A, B$ and $E$ will represent the quantum sytems of Alice, Bob and Eve.

There are several possibilities for Bob's measurement, specifically if $m$ is odd. 
We will limit Bob's apparatus to single-photon detector based set-ups,
similar to the one used in the BB84 and SARG04 protocols. 
This  limitation will prevent Alice and Bob to extract all the 
information allowed by the Holevo bound \cite{Holevo73, Holevo98, SchumacherWestmoreland97} 
or to use continuous-variable detection set-up like the one used in the LG09 protocol 
\cite{LeverrierGrangier09}.

Since Bob's measurement is based on single-photon detectors, Alice and Bob
need to postselect-away the event when Bob has received no photon \ie when Bob's 
detectors have not clicked. This can be done by one-way classical communication from 
Bob to Alice. The kept events typically constitute a fraction $Y=1-e^{-T\mu}=T\mu + \mathcal O(T^2\mu^2)$ of the sent pulses if the sent WCPs have a mean photon number of $\mu$, and are sent through a channel of transmission $T$.

When Bob receives a single photon, he performs the POVM 
$\left\{\tfrac2m\ketbra{y\theta_m+\pi}\right\}_{y\in\llbracket0,m-1\rrbracket}$.
The $\pi$ dephasing  does not change anything if $m$ is even, 
but increases the mutual information $I(X{:}Y)$ between 
Alice and Bob when $m$ is odd. 
In particular, it ensures that, for any state sent by Alice, one outcome ($y=x$) 
of Bob's measurement is impossible.

\subsection{Keyrate without eavesdropping}
\label{cleanKeyrate}

If Alice sends perfect single photon pulses, the lack of errors guarantees a 
perfect secrecy of  $I(X{:}Y|m)$ bits of the  key, where  $I(X{:}Y|m)$ is 
the mutual information between Alice and Bob's data.
This quantity is easily derived from the conditional probabilities 
\begin{equation} \label{eq:CondProb}
  \Proba{y|x, m}=\tfrac1m\left(1-\cos(y-x)\theta_ m\right).
\end{equation}
 The equiprobability of the $m$ states encoded by Alice, 
 ensures the equiprobability of the $m$ possible values measured
 by bob. The Shannon entropy of Bob's measurement is  therefore
\begin{equation}
H(Y|m)=\log m  
\end{equation}
and then 
\begin{align}
  H(Y|X, m)&=\log m -\frac1m\sum_{k=0}^{m-1}
      (1 -\cos k\theta_m)\log(1 -\cos k\theta_m);\\ 
  I(X{:}Y|m)&= H(Y|m) -  H(Y|X,m) \\
    &= \frac1m\sum_{k=0}^{m-1}
      (1 -\cos k\theta_m)\log(1 -\cos k\theta_m). \label{mutualAB}
\end{align}

The above sum can be seen as the $m^{\text{th}}$ Riemann sum of 
$(1 -\cos k\theta)\log(1 -\cos k\theta)$ taken with $m$ samples. 
It therefore decreases slightly with $m$, 
from $\log\tfrac32=0.5850$ bits for $m=3$ 
to $\tfrac1{2\pi}\int(1 -\cos k\theta)\log(1 -\cos k\theta)d\theta=0.4427$ bits
in the continuous limit $m\rightarrow\infty$.
For $m=4$, we have $I(X{:}Y|m=4)=\frac12\log2$, 
\ie the same rate as BB84.

When Bob receives more than one 
photon, several detectors can click. This gives him more information than single 
clicks, so neglecting this case, as done above, is pessimistic. 
This corresponds to Bob randomly choosing between the various detection results.

In a reverse reconciliation (RR) scheme \cite{GrosshansGrangier02-Proc, GVAWBCG03},
Alice and Bob can share a common key of length $I(X{:}Y)$ provided Bob sends to 
Alice at least $H(Y|X)$ bits of 
side-information. For example, when $m=4$, 
Bob needs to send 1.5 bits per pulse. This can be done by revealing his 
measurement basis (1 bit/pulse) and using the syndrome of a good erasure correcting
 code (see \eg \cite[Chapter 50]{MacKay}),
 which will be slightly longer than $\frac12$ bit/pulse. 
Indeed, when Bob has revealed his basis measurements, Alice knows which bits of $Y$ are not correlated with $X$
and this corresponds to an erasure channel of rate $\frac12$.
Note that $I(X{:}Y|m=4)$ is the same as the mutual information 
but that Alice and Bob need only 1.5~bits of one-way communication to extract
the secret key, instead of 2 bits of two-way communication.
Their use of erasure correcting codes
instead of interactively throwing some bits away is at the heart of
the resistance of this protocol against PNS attacks: on 2-photon-pulses, Eve 
can keep a copy of the pulse sent by Alice, and, even if she knows the basis of 
Bob's measurement, she ignores  whether Alice sent a state in the right basis 
or not. Therefore, in this case, Eve's
 measurement has at best a $25\%$ error-rate, 
giving her at most $h(\tfrac14)=0.1887$ bits of information
--- where $h(\cdot)$ is the binary entropy ---
while Alice still has half a bit. The net keyrate of 2-photon-pulses is then 0.3113 bits.
In BB84, on the contrary, Alice reveals her basis choice,
living her on equal footing with Eve for 2-photon-pulses.

Note that when $m$ is even, the above idea for the reconciliation can be
generalized \ie Bob reveals $\log m -\log2$ bits for the basis $\left\{y \left[\tfrac m2\right]\right\}_{y=0}^{m/2-1}$ 
where the square bracket $[\cdots]$ is a short-hand for 
``modulo $\cdots$''.
Bob then uses the appropriate error correcting code for the remaining information. We are then in 
a situation where Alice has different known error rates 
$\tfrac12\left\{1-\cos\left(\left(x-y \left[\tfrac m2\right]\right)\theta_m\right)\right\}$ 
for different bits while Eve only sees the average error rate.
The following paragraphs will study the above statements more formally, in
the asymptotic and error-less regime.

\section{Two families of eavesdropping attacks}
\label{secEve}
If Alice sends perfect single photon pulses, the lack of errors guarantees a perfect secrecy of  
$I(X{:}Y|n=1)$ bits of the  key.
However, if Alice uses weak coherent pulses (WCP) some attacks become possible 
without introducing errors. Eve can either perform an intercept resend with unambiguous state discrimination attack (IRUD)
 \cite{Chefles-98, felix-01}  (see section \ref{subsecIRUD}), at the price of blocking some pulses, 
or photon number splitting attacks (PNS) \cite{huttner-95, yuen-96, brassard-00} 
(see section \ref{subsecPNS}), as well as a combination of the two attacks.

%
%
In any case, since Alice's pulses are phase randomized,
Eve's optimal attack can be described by a quantum 
non-demolition measurement of the photon number 
$n$  of Alice's pulse followed by an action which depends on $n$
 \cite{Luetkenhaus00}.

 \subsection{Some useful notations} 
 \label{appendix-rewriting-states} 

In order  to simplify the analysis, the state \eqref{statextheta} 
of $n$ photons polarized in the direction $x\theta_m$ can be written

\begin{align}
\ket{x,n,m}&=\ket{x\theta_m}^{\otimes n}
  =2^{-\frac n2}(\ket0+e^{i x \theta_m}\ket1)^{\otimes n}\\
  &=2^{-\frac n2}\sum_{b=0}^{2^n-1}e^{i\Hamming b x \theta_m}\ket b, 
\end{align}
where $\ket b$ is the tensorial binary development of 
$b\in\llbracket0,2^n-1\rrbracket$,  
and $\Hamming b$ its Hamming 
weight, \ie  the number of bits set to 1 in the binary development of $b$. 
Note that all terms with the same Hamming weight $w$ modulo $m$ have the 
same phase prefactor $e^{iw\theta_m}=e^{i(w[m])\theta_m}$. 
These $\binom{n}{w[m]}$ vectors are orthogonal. We have defined
\begin{equation}
\binom{n}{w[m]}\isbydef
    \sum_{d=0}^\infty    \binom{n}{w+dm},
\end{equation}
where $\binom{n}{w+dm} = 0$ for $w + dm > n$.
Let's define, for each $w\in\llbracket0,m-1\rrbracket$,
\begin{equation}
  \ket{w[m]}_n\isbydef
    \frac1{\sqrt{\binom{n}{w[m]}}}
    \sum_{\substack{b\in\llbracket0,2^n-1\rrbracket\\
      b\equiv w[m]}}
    \ket{b}.
\end{equation}
We can then rewrite the state $\ket{x,n,m}$ as
\begin{equation}
  \ket{x,n,m}=
    2^{-\frac n2}
    \sum_{w=0}^{m-1}e^{i w x \theta_m}\textstyle{\sqrt{\binom{n}{w[m]}}}\ket{w[m]}_n.
  \label{Aeq:ketxnm}
\end{equation}

For the following discussion it  will also be useful to mentally 
group the pulses with the same number of photons $n$. For each of those groups we can then define $b_{n}$  as the fraction of blocked pulses, $u_{n}$ as the fraction of pulses on which an unambiguous discrimination is successful, and $p_{n}$ as the fraction where a PNS attack is performed.  For each value of $n$ it is also useful to define the specific conditional probabilities $Y_{n}$ (photon-number dependent yield), defined as the probabilities for Bob to see his detector clicking when Alice has sent a $n$-photon-state:
\begin{equation}
Y_{n} = (1- b_{n}) \label{n-dep-yields}.
\end{equation}
If for each pulse we represent with $P_{n|\mu}$ the probability of having $n$ photons knowing that the average photon number is $\mu$, we have
\begin{equation}
\label{yield-from-mu}
 Y=\sum_{n=0}^{\infty}P_{n|\mu}Y_n.
\end{equation}

\subsection{IRUD attacks}
\label{subsecIRUD}
As shown in equation \eqref{Aeq:ketxnm},
the state of an $n$-photon-pulse sent by Alice can be written
\begin{equation}
  \ket{x,n,m}=
    2^{-\frac n2}
    \sum_{w=0}^{m-1}e^{i w x \theta_m}\textstyle{\sqrt{\binom{n}{w[m]}}}\ket{w[m]}_n.
  \label{eq:ketxnm}
\end{equation}

If Eve makes an IRUD attack, the success probability of her unambiguous measurement
 is given in \cite{CheflesBarnett98} as
\begin{align} \label{Pdelta}
  \Proba{\Delta|m,n}&=2^{-n}m\min_{w\in\llbracket0,m-1\rrbracket}
    \textstyle{\binom{n}{w[m]}}.
\end{align}
This probability is not null iff $n\ge m-1$, and its value 
increases each time $n$ increases by 2. Its first nonzero value is $2^{-m+1}m$ for
$n\in\{m+1,m+2\}$.
Since she does not want to introduce errors, Eve has to block the pulses on which an IRUD attack has failed.
She can then resend (with no errors) a fraction $u_n$ 
of the original pulses as big as
\begin{equation}
  u_n=\min\left(\frac{\Proba{\Delta|n,m}}{1-\Proba{\Delta|n,m}}b_n; 1-b_{n}\right).
\end{equation}
In other words, she can intercept and resend 
$\frac{\Proba{\Delta|n,m}}{1-\Proba{\Delta|n,m}}$ pulses for each pulse she 
blocks, without introducing any error.
For example, for $m=4$ and $n=3$, Eve can perform an unambiguous state 
discrimination with a success probability $\Proba{\Delta|n=3,m=4}=\frac12$. This
means that, for each 3-photon-pulse Eve has blocked, there is another she has 
unambiguously discriminated and resent to Bob without error.

On remaining $p_n=1-b_n-u_n$ pulses, she can perform a PNS attack, keeping $n-1$ 
photons
and transmitting the remaining one unperturbed to Bob.
We have
\begin{equation}
p_n=  \left[  \tfrac{1-\Proba{\Delta|n,m}-b_n}{1-\Proba{\Delta|n,m}}  \right]_{+}
\end{equation}
where $[\cdots]_+$ is a shorthand for $\max(\cdots;0)$. 


\subsection{PNS attack}
\label{subsecPNS}

Since one can construct a Markov chain 
$Y\leftrightarrow X \rightarrow \ket{x,n-1,m}$,
and since the latter is the state held by Eve when she performs a PNS attack,
$\chi(Y{:}E|n,\text{PNS})< I(Y{:}X)$. 
The inequality is strict because the last transition is not reversible.
In other words, PNS attacks without IRUD
can never reduce the 
net RR-keyrate $K_n=S(Y{:}X)-S(Y{:}E|n,\text{PNS})$ to $0$, 
contrarily to the BB84 protocol.

In order to compute the efficiency of the PNS-attack, one needs to compute 
the density matrices associated with $n$-photon-pulses.
The density matrix corresponding to the state defined in \eqref{eq:ketxnm} is:
\begin{align}
    \ketbra{x,n,m}=
      2^{-n}
        \sum_{w,w'=0}^{m-1}e^{i(w-w')x\theta_m}\textstyle{\sqrt{\binom{n}{w[m]}\binom{n}{w'[m]}}}
      \notag\\\times
        \ket{w[m]}\bra{w'[m]}
      \\
      =\sum_{D=1-m}^{m-1}
        e^{iDx\theta_m}\M_{D,m,n},
\end{align}
  where we have defined, 
  for any integer $D\in\llbracket1-m,m-1\rrbracket$, the
  (shifted) $m\times m$ diagonal matrix
  \begin{equation}
  \M_{D,m,n}
    \isbydef
      2^{-n}\sum_{w=0}^{m-1}
      \textstyle{\sqrt{\binom{n}{w[m]}\binom{n}{w+D[m]}}}
      \ket{w[m]}\bra{w+D[m]}.
  \end{equation}

Let $\rho_{n,m}$ be the generic $n$-photon-state sent by Alice.
One has then
\begin{equation}
  \rho_{n,m}
    =\sum_{x=0}^{m-1}\tfrac1m\ketbra{x,n,m}
    =\M_{0,m,n}.
\end{equation}

When Bob measures $Y=y$, and
Eve keeps $n$ photons, her state conditioned on Bob's measurement is given by
\begin{align}
  \rho_{y,n,m}&=\sum_{x=0}^{m-1}\tfrac1m(1-\cos(x-y)\theta_m)\ketbra{x,n}\\
    &=\rho_{n,m}
    -\tfrac1m\sum_{x=0}^{m-1}\cos(x-y)\theta_m
      \sum_{D=1-m}^{m-1}e^{ixD\theta_m}\M_{D}\nonumber\\
    &=\M_{0}
    -\sum_{D=1-m}^{m-1}\frac{\M_{D}}{2m} 
      \sum_{x=0}^{m-1}\left\{e^{i\theta_m(x(D+1)-y)}
          +e^{i\theta_m(x(D-1)+y)}\right\}\nonumber\\
    &=\M_{0}
      -\tfrac{e^{-iy\theta_m}}2(\M_{m-1}+\M_{-1})
      -\tfrac{e^{iy\theta_m}}2(\M_{-m+1}+\M_1).
\end{align}  
Note that in the above equations, the indices $m$ and $n$ have been omitted for 
$\M_D$ for the sake of simplification.

The Holevo limit of the information Eve can gather on Bob's measurement through
 a collective PNS attack is
\begin{align}
   \chi(Y{:}E|n,\text{PNS})
     &\isbydef S(E|n,\text{PNS}) - S(E|Y,n,\text{PNS})\\
     &=S(\rho_{n-1,m})-S(\rho_{y,n-1,m}) \label{mutualEB}.
 \end{align}
 These entropies are easily computed numerically and
 decrease slowly with $n$.
 
 Since they are independent of $m$ iff
$n\le m-1$, the corresponding $S(Y{:}E)$ will also be identical 
in this case. In other words, the information leaked to Eve in $m$-state protocols 
are identical to the 
continuous $m\rightarrow\infty$ limit for $n$-photon-pulses 
when $n\le m-2$, and
the only difference at $n=m-1$ comes from the IRUD attack.

\section{Global eavesdropping strategies}
\label{secEavesStrat}

In this section we  study how Eve can exploit the imperfect transmission of the channel and the multiphoton pulses, and optimally combine the IRUD and PNS attacks described above to decrease Alice's and Bob's keyrate $K$. 

In subsection \ref{sub-section-IIIA} we  compute  the total  keyrate as a function of the photon number distribution $P_{n|\mu}$ and the photon-number-dependent yield $Y_{n}$.  This  expression of the total keyrate will be then written explicitly for the two cases discussed   in the two following subsections.  

In subsection \ref{subsecKnoDSP} we  study the 
optimal strategy for Eve when Alice and Bob use coherent states of fixes average 
photon number $\mu$.

In subsection \ref{subsecKDSP} we  consider the case where a \emph{decoy-states protocol} (DSP) is applied \cite{Hwang-03, LoMaChen-05, Wang-05}. We will see how the knowledge of Alice and Bob of the yields $Y_{n}$ results in more constraints for Eve.

Finally, in subsection \ref{comparisons-section} we compare the keyrates obtained for the sifting-less protocol to the keyrates of BB84.

\subsection{Total keyrate as function of $\mu$ and $\{Y_{n}\}$}
\label{sub-section-IIIA}

 In this first subsection we will use the IRUD and PNS attacks introduced in section \ref{secEve} to compute a general expression of the total keyrate as a sum over $n$ of   contributions due to each (group of pulses with a given) number of photons. This total keyrate will be a function of the average number of photons $\mu$ and the photon-number dependent yields $\{Y_{n}\}$. As we have seen, the pulses attacked with IRUD carry no secret information, since they are either blocked or perfectly known by Eve. So the  pulses with a non-null contribution to the keyrate are those on which only a PNS attack has been performed.

\begin{figure}
  \includegraphics[width=\columnwidth]{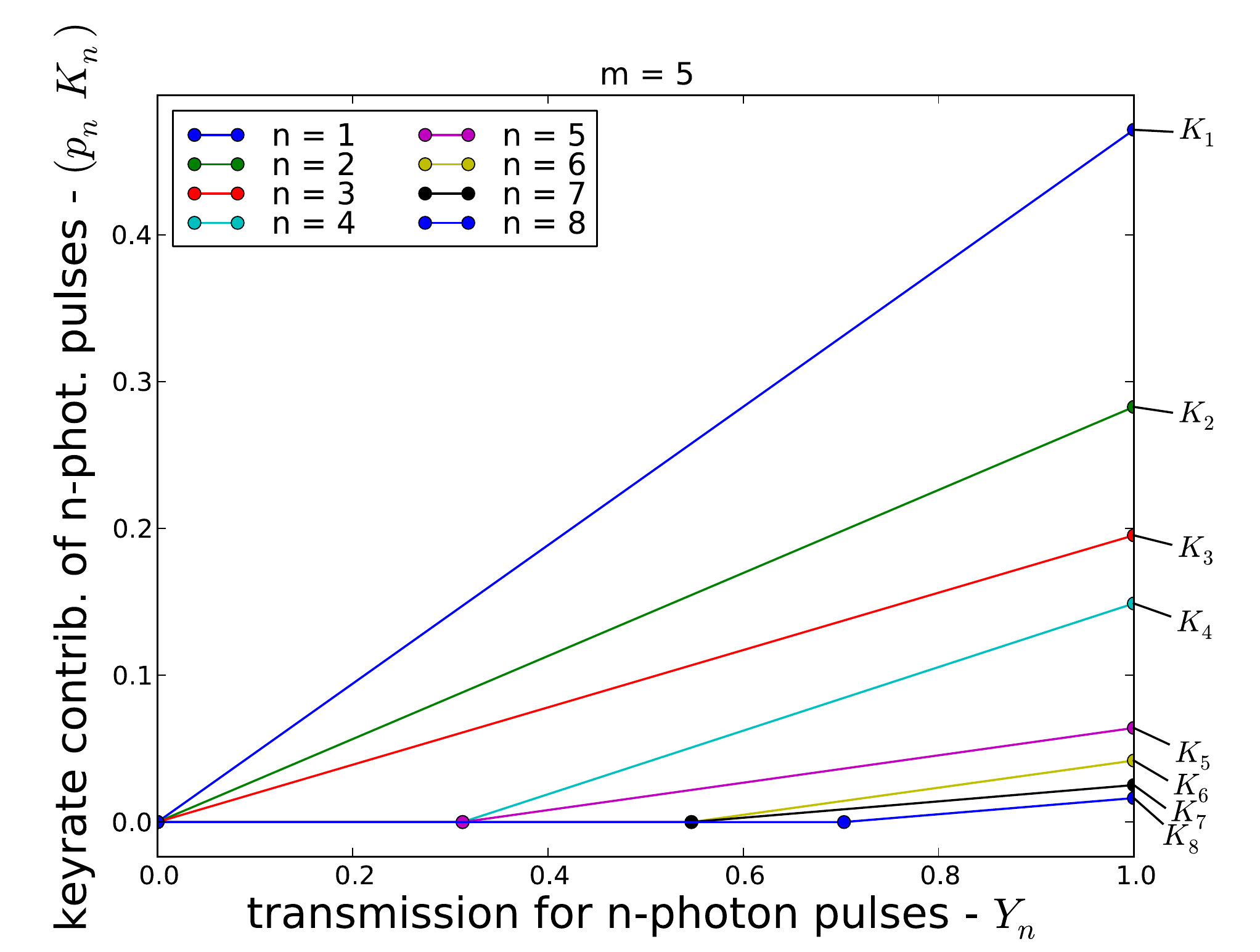}
  \caption{\label{Ktilde-n} (color online) Plots of $(p_{n} K_{n})$ for $m=5$ and for several values of $n$, as function of $Y_n$. These functions can be seen as the terms of the sum in formula \eqref{Ktot-compact}. We see how these are qualitatively similar linear functions, with a threshold at $Y_{n} = \Pdelta$ where it becomes non-zero, and a different slope $K^{\text{marg}}_{n,m}$ for each value of $n$ and $m$. The value of the threshold $\Pdelta$ is  zero for $n < m-1$.}
\end{figure}

The keyrate contribution of $n$-photon-pulses for a given yield $Y_{n}$ depends on two quantities: the probability of success $\Pdelta$ of the \irud attack, described in \eqref{Pdelta}, and the maximal keyrate contribution $K_n$ for a lossless  channel ($Y_n=1$):
\begin{equation} \label{Kn}
K_n\isbydef S(X{:}Y)-S(Y{:}E|n,\text{PNS})
\end{equation}
which is obtained subtracting the Holevo quantity between Bob and Eve \eqref{mutualEB} from the mutual information between  Alice and Bob \eqref{mutualAB}. We can write: 
\begin{equation} \label{ktot-with-pn}
K(\{Y_{n}\},\mu) =\sum_{n=1}^{\infty} P_{n|\mu} p_n K_n,
\end{equation}
where $P_{n|\mu}$ is the probability to have $n$ photons in a WCP and $p_{n}$ has been defined at the beginning of section \ref{secEve} as the fraction of 
$n$-photon-pulses on which Eve performs PNS attacks.
Then, using the relations $p_{n} = 1-\frac{u_{n}}{\Pdelta} = 1-\frac{b_{n}}{1-\Pdelta}$ and $b_{n} = (1-Y_{n})$ we have:
\begin{equation} \label{Ktot}
  K(\{Y_{n}\},\mu) =\sum_{n=1}^{\infty} P_{n|\mu}  K_n -
    \sum_{n=1}^{\infty}P_{n|\mu}  (1-Y_n)\frac{K_n}{1-\Pdelta} 
\end{equation} 
For given values of $m$ and $n$, the product  $(p_{n} K_{n})$ describes the keyrate contribution of 
$n$-photon-pulses. 
Its value varies with the channel yield $Y_n$. 
Figure \ref{Ktilde-n}  shows a plot of $(p_{n} K_{n})$ for $m=5$ and several values of $n$.  
 These functions are constantly zero for $Y_n \leq \Pdelta$, 
and then rise linearly up to the maximal value $K_{n}$ reached for $Y_n=1$.  

\begin{figure}
  \includegraphics[width=\columnwidth]{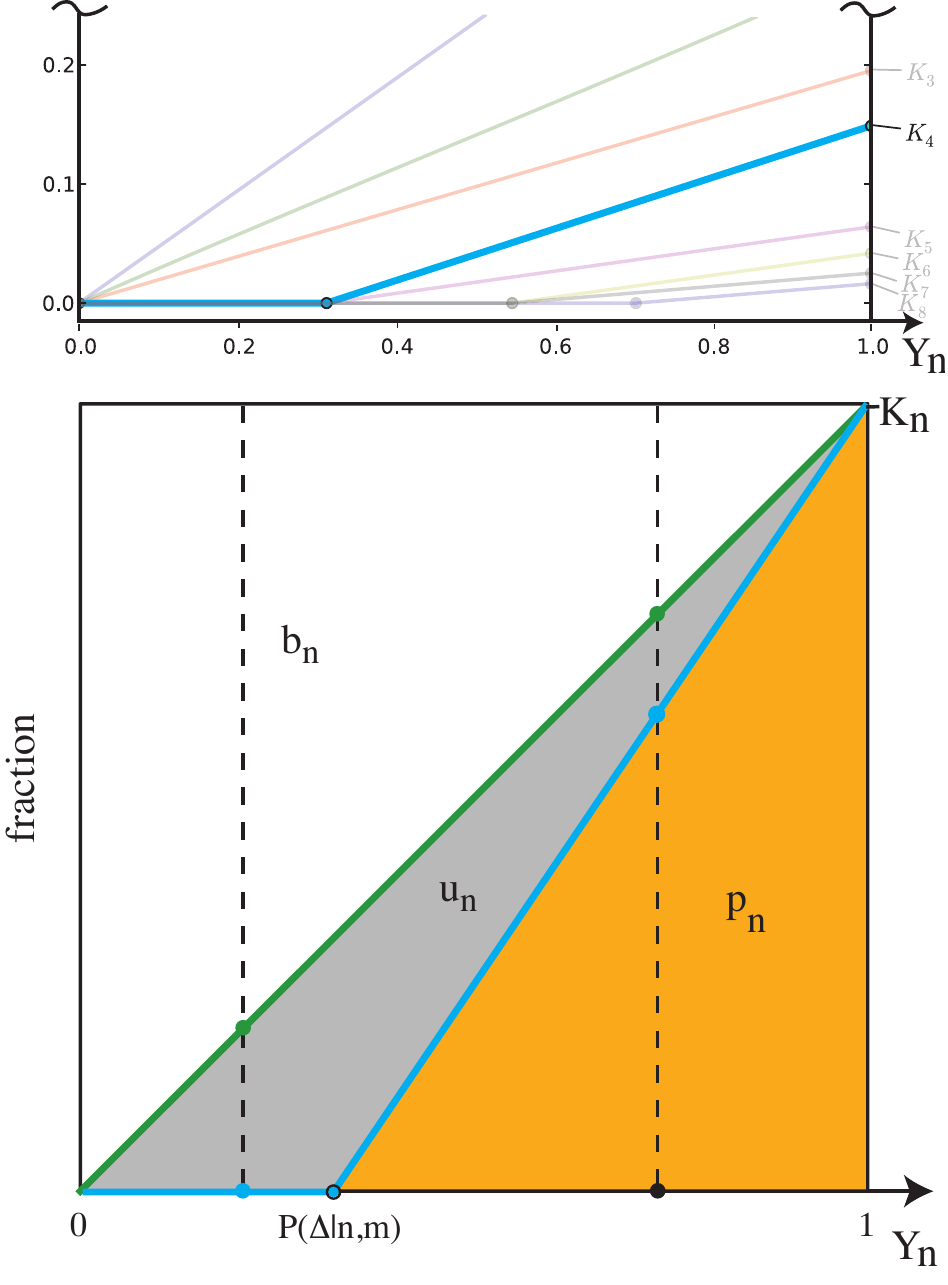}
  \caption{\label{plotWsquare} (color online) In the upper part of this figure we report part of the plot in figure \ref{Ktilde-n}, where we have highlighted one curve out of the others (the one for $n=4$). In the lower part we have a square with two curves which delimit three regions. The horizontal axis of the square is the same as the upper plot, \ie $Y_n$, while the vertical axis represents fractional values, the whole width of the side being ``normalized'' to $1$. The blue line is the curve from the upper part plot, rescaled to $K_{n}=1$, while the green line is a 0.5 slope line which reports on the vertical axis the fraction $Y_n$. The three regions represent the fractions of pulses in a group with fixed number of photons: $b_{n} =$ blocked pulses (\irud failure), $u_{n}$ = pulses on which the \irud attack is successful, $p_{n}$ = fraction on which the \pns attack is performed.}
\end{figure}

The slopes
\begin{equation} \label{marg-keyrate}
K_{n, m}^{\text{marg}}  \isbydef \frac{\partial(p_{n}K_{n})}{\partial Y_{n}} = \frac{K_n}{1-\Pdelta}.
\end{equation}
define the marginal keyrates $K^{\text{marg}}_{n, m}$.  These marginal keyrates express the contribution to the total keyrate brought by each 
$n$-photon-pulse, when $Y_{n}\geq\Pdelta$. 
As  shown in figure \ref{rates-figure}, 
the value of the slopes $K_{n, m}^{\text{marg}}$, is often decreasing with $n$, but not always.
We can now rewrite the total keyrate \eqref{Ktot} in a more compact form as:
\begin{equation} \label{Ktot-compact}
K = \sum_{n} \left[Y_{n} -  \Pdelta  \right]_{+} P_{n|\mu} K^{\text{marg}}_{n, m}
\end{equation}
where  again, the ``$+$'' subscript means that the value of the square bracket is 
is set to zero if the contained expression is negative
and is otherwise kept positive.

\subsection{Keyrate  \emph{without} decoy-states protocol:\\ the ``budget strategy''} 
\label{subsecKnoDSP}

When Alice and Bob do not use a DSP, they do not know the different $Y_n$, and  they only measure the global transmission yield $Y$.

Eve is then free to optimize her attack considering only this constraint of the total yield of the channel $Y$, and limit the total number of \irud attacks, so that the total fraction of blocked pulses
  is $\sum_{n} b_{n} \leq 1-Y$.

In other words, Eve has a total budget of $1-Y$ pulses to block. When she blocks 
a $n$-photon-pulse she decreases the keyrate by 
$K_{n, m}^{\text{marg}}$ if $Y_{n}>\Pdelta$ but doesn't change it if $Y_{n}\leq\Pdelta$. Her best strategy is therefore to 
wisely spend her budget, 
attacking the pulses with an higher marginal keyrate \eqref{marg-keyrate} in pritority to the ones with a lower marginal keyrate. 

The first group of pulses on which Eve can apply IRUD, is usually the one with  $m-1$ number of photons.  Then,  for each group Eve behaves differently, 
depending whether the value of $Y_n$ is lower or higher than $\Pdelta$. 
If $Y_n\leq \Pdelta$ (see left vertical dashed line in lower part of figure \ref{plotWsquare}) she applies the \irud attack to all the pulses of the group, given that she has enough ``blocking budget'' to spend. She stops performing IRUD attacks when the total amount of blocked pulses reaches the channel's losses value $(1-Y)$. Eve applies the \pns attack to all the other $n$-photon-pulses.

On the other hand, if $Y_n\geq\Pdelta$ for a group of $n$-photon-pulses  (see right vertical dashed line in lower part of figure \ref{plotWsquare}),
she applies \irud only to a fraction $1-(p_{n} K_{n})$ of the group of $n$-photon-pulses (see first two upper parts of the vertical line, in the $b_{n}$ and $u_{n}$ regions), and then applies \pns attack to the remaining pulses of the group. 

If Eve has not enough blocking budget to perform an \irud attack, she will perform a \pns attack on all the remaining pulses.

\begin{figure}
  \includegraphics[width=\columnwidth]{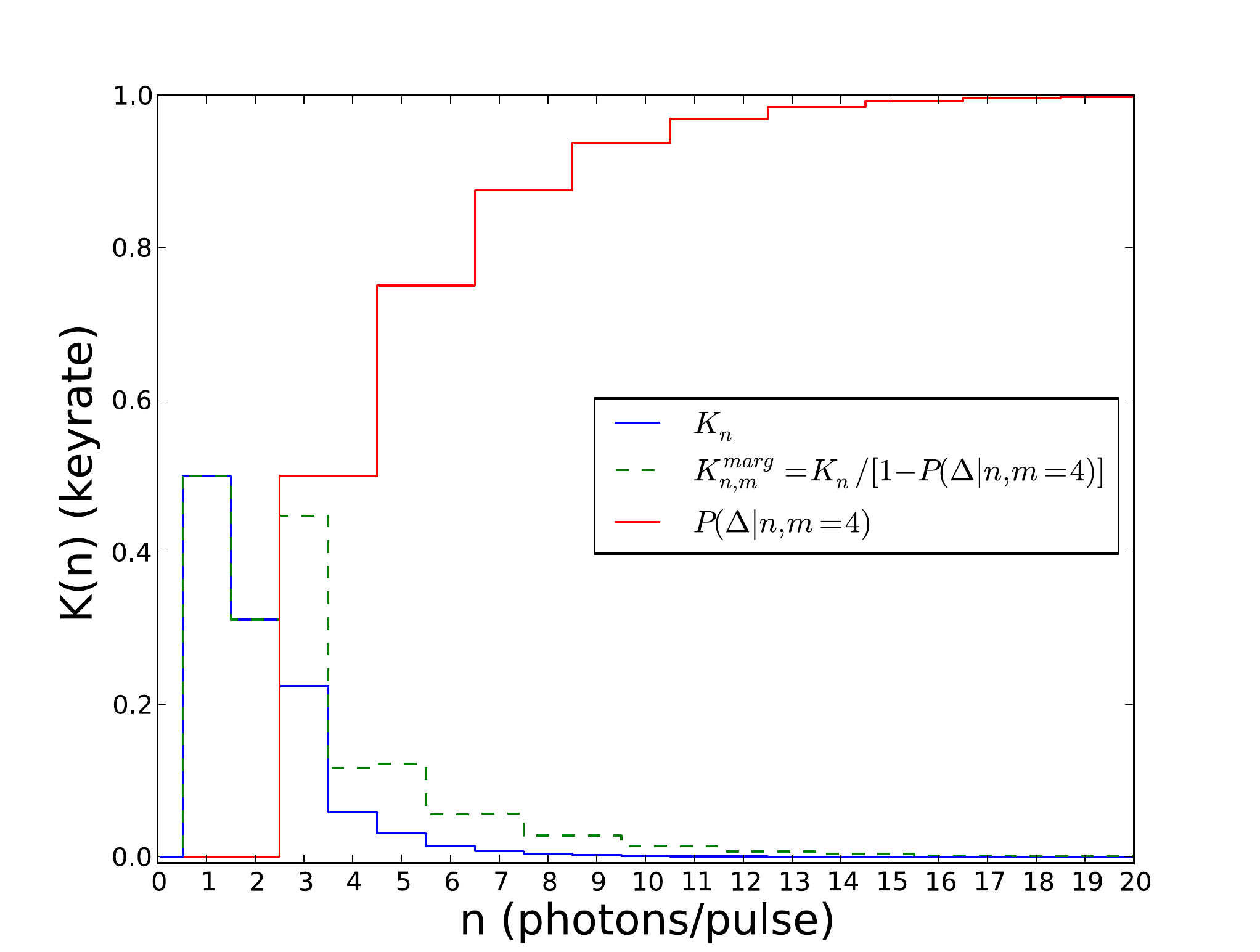}
  \caption{\label{rates-figure}
  (color online) Plot of the behaviour of several quantities linked to the contribution to the keyrate, as function of the number of photons, for $m=4$ protocol states. The green dashed line represents $K_{n, m}^{\text{marg}}$ (see equation \eqref{marg-keyrate}) which is the slope of the non-zero part of $(p_{n} K_{n})$ (see figure \ref{Ktilde-n}) contribution of a single pulse to the keyrate. Red line represents $\mathcal P(\Delta |n,m=4)$ (see equation \eqref{Pdelta}), the probability of success of IRUD attacks, which appears at the denominator of $K_{n, m}^{\text{marg}}$, and blue line represents the maximal keyrate $K_{n}$, which appears at the numerator of $K_{n, m}^{\text{marg}}$, and is described in equation \eqref{Kn}. From these plots we can see how $K_{n, m}^{\text{marg}}$ is not monotone with respect to $n$, and for example is bigger for $n=3$ than for $n=2$ }
\end{figure}

The behavior of the effective keyrate with respect to the number of photons is described in figure \ref{rates-figure}. As an example it shows how going from 2- to 3-photon-pulses the contribution to the keyrate per pulse increases, making more convenient for Eve to first apply IRUD to the pulses with 3 photons, and only then to those with 2.

\begin{figure}
  \includegraphics[width=\columnwidth]{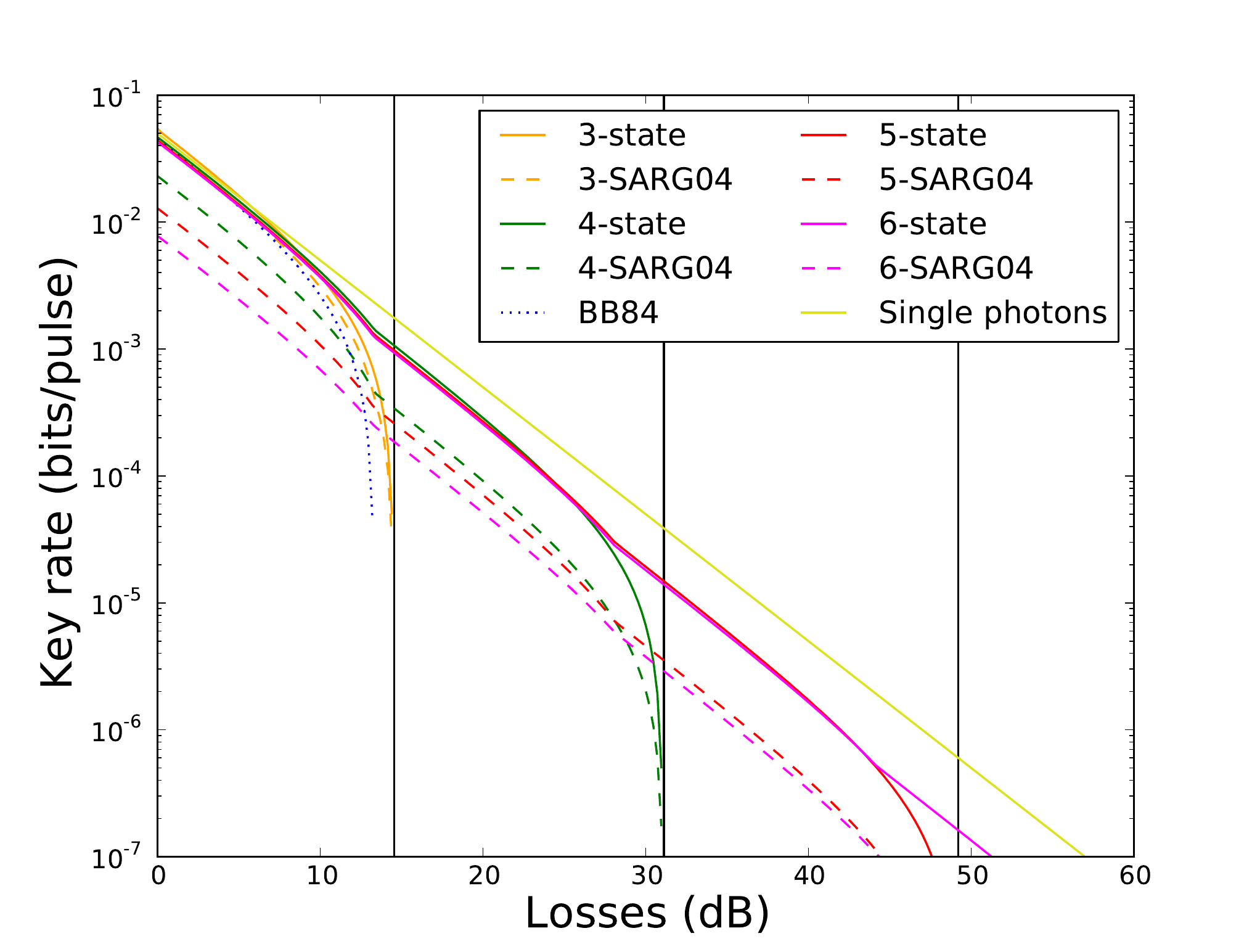}
  \caption{\label{FigKeyRates}
  (color online) Keyrates of the $m$-states protocols compared to $m$-state SARG04 and BB84
  with WCPs for $\mu=0.1$ and $m\in\llbracket3,6\rrbracket$. 
  The vertical lines represent the values of $T_c$ given by \eqref{eqTc}.}
\end{figure}

The keyrates for fixed $\mu=0.1$ are plotted in figure \ref{FigKeyRates}.

The net keyrate is $0$ when all the lost pulses could be explained  
by IRUD attacks, \ie when $\forall n, p_n=0$. 
Let $T_c$ be the critical transmission below which our protocol ceases to work.
At transmission $T_c$, all the $1-e^{-T_c\mu}\simeq T_c\mu$ transmitted pulses correspond to 
the $e^{-\mu}\sum_{n=0}^{\infty}\frac{\mu^n}{n!}\Proba{\Delta|n,m}$
successful IRUD attacks. 
We have then
\begin{equation}
  T_c=-\tfrac1\mu\ln\left[1-e^{-\mu}\sum_{n=0}^{\infty}\tfrac{\mu^n}{n!}\Proba{\Delta|n,m}\right]
  .
  \label{eqTc}
\end{equation}

If $\mu\ll1$, the sum is dominated by the first non-zero term, corresponding to
$n=m-1$ and we have

\begin{align}
  T_c \mu&\simeq\frac{\mu^{m-1}}{m-1!}\frac m{2^{m-1}}\\
  T_c&\simeq\frac{\mu^{m-2}}{m-1!}\frac m{2^{m-1}}\\
  T_c&\simeq\frac{m}{2\cdot m-1!}\left(\frac\mu2\right)^{m-2}.
\end{align}
We essentially have $T_c\propto \mu^{m-2}$, showing the exponentially increasing 
robustness  of the protocol for increasing $m$. 
This dependency is 
the same as SARG04,  
but not as BB84,
where $T_c\simeq\frac\mu2$.

\begin{figure}
  \includegraphics[width=\columnwidth]{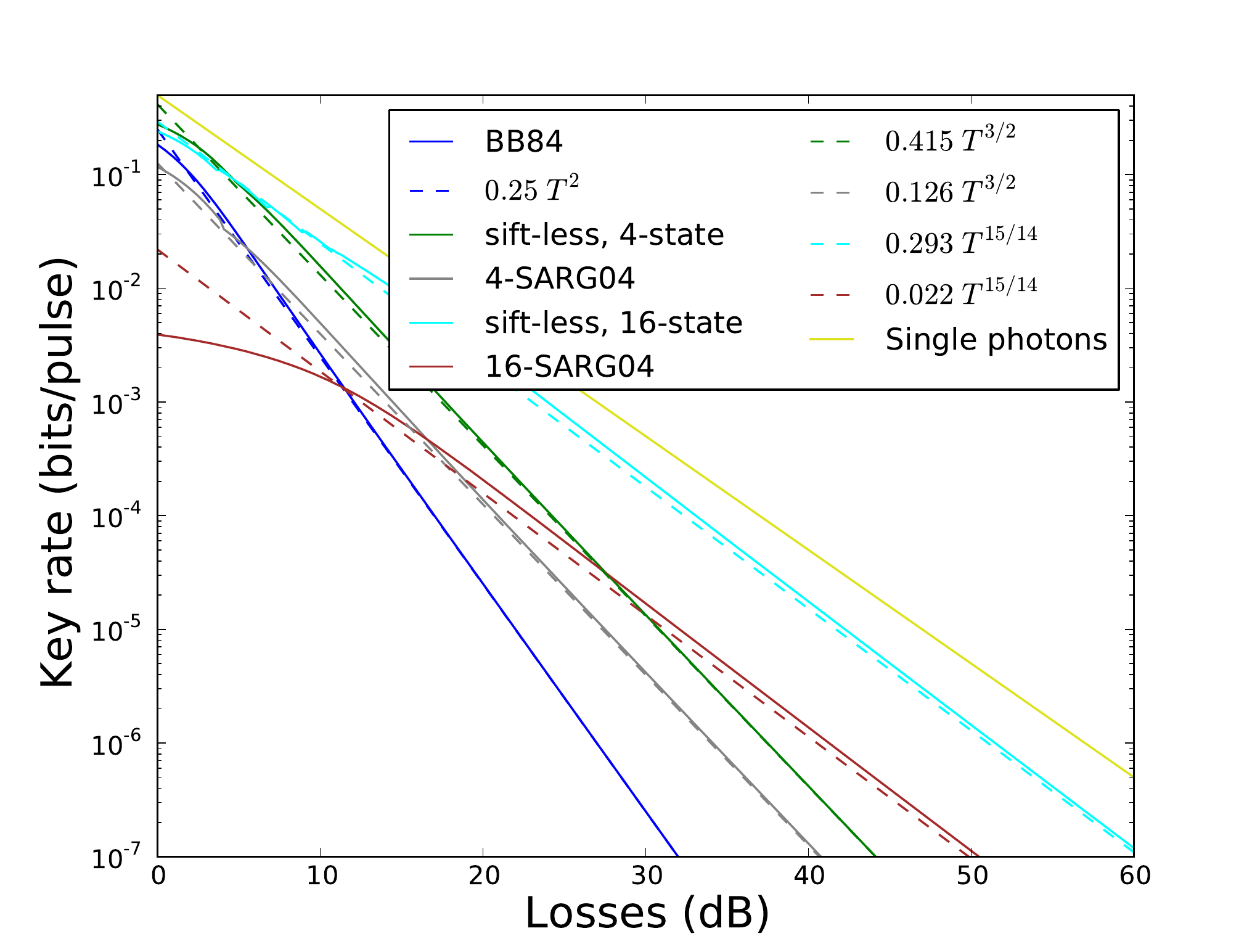}
  \caption{\label{FigOptKeyRates}
  (color online) Keyrates with optimized $\mu$ for BB84, the $m$-states protocol, 
  $m$-state for $m=4$ and $m=16$. Dashed lines represent the approximations described in equation \eqref{k-opt}. 
  }
\end{figure}

One can also numerically optimize $\mu$ for each value of the transmission $T$,
as shown in figure \ref{FigOptKeyRates}.

If the optimal keyrate is achieved close to $T_c$, we have,
for $\mu\ll1$, 
\begin{align}
  K&\simeq K'_{m-1}\left(T\mu - \Proba{\Delta|m-1,m}\tfrac{\mu^{m-1}}{m-1!}\right)\\
\intertext{with $K'_{m-1}$ being the $(m-1)^{\text{th}}$ value of the 
$\frac{K_n}{1-\Proba{\Delta,n,m}}$ coefficients in decreasing order. 
Optimizing this quantity for $\mu$ is straightforward and gives
}
\mu_{\text{opt}}&\simeq2 \left(\tfrac{2\cdot m-2!}{m}\right)^{\frac1{m-2}}  T^{\frac1{m-2}}\\
  K_{\text{opt}}&\simeq
      K'_{m-1}
      \tfrac{2}{m-1} 
      \left(\tfrac{2\cdot m-2!}{m}\right)^{\frac1{m-2}}
      T^{1+\frac1{m-2}} \label{k-opt}
\end{align}
\ie the keyrate essentially varies as  
$K\propto T^{1+\frac1{m-2}}$ with a prefactor which slowly 
decreases with $m$. This approximation seems in agreement
with numerical results, at least for reasonably low $m$ (below 16).
The bigger $m$ is, the closer one is to the ideal single-photon case,
where $K=\frac T2\log 2$.

\subsection{Keyrate,  \emph{with} decoy-states protocol}
\label{subsecKDSP}

%
The DSPs are  essentially designed to deprieve Eve from  
the freedom to play with the yields $\{Y_{n}\}$ \cite{Hwang-03, LoMaChen-05, Wang-05}. 
More precisely, in a DSP, Alice randomly intermixes the pulses with nominal mean photon number $\mu_0$ used for the normal protocol, 
with decoy pulses with $D$ different mean photon number $\mu_d$, with $d\in\llbracket1,D\rrbracket$.
After Bob has performed his measurement, Alice reveals the intensity used for each pulse and 
Bob says whether he has measured a click or not. They can then compute the yield $Y_{\mu_d}$ of WCP of mean photon number $\mu_d$ and
the equation \eqref{yield-from-mu} becomes a system of $D+1$ linear equations with infinitely 
many unknowns:
\begin{equation}
  Y_{\mu_d}=\sum_{n=0}^{\infty}P_{n|\mu_d}Y_n.
\end{equation}
By setting $Y_n=0$ for $n>D$, 
Alice and Bob
can reduce the number of unknowns to $D+1$ and solve this system of equation for the 
yields $\{Y_n\}_{n\in\llbracket0,D\rrbracket}$
This correspond to pessimistically assume that the pulses with $n>D$ do not contribute to the yield,
but they can do that safely  only bay assuming that these pulses still leak all the information 
$I(X,Y)$ to Eve. 
Since we are interested here in the asymptotic limit of many pulses, where the finite size-effects are negligible, the statistic gathering only needs a small fraction of decoy pulses, no matter how big (but finite) $D$ is. Therefore, we will assume from now on that this DSP procedure is ideal and
gives Alice and Bob  the yields $Y_n$ of the channel for each photon number $n$.

 Indeed, if a \dsp   is performed,  the ``budgeting strategy''  described in the previous subsection \ref{subsecKnoDSP} has to be modified, and the security analysis goes back to the general discussion given in subsection \ref{sub-section-IIIA}. In this case Eve can not take into account only the total transmission of the channel, so she can not perform any optimization, and has to apply the \irud attacks  considering each $n$-photon-group separately. 
Moreover, as before  Eve has also to consider whether  $Y_n\leq \Pdelta$ or $Y_n\geq \Pdelta$. In the first case she applies  \irud attack to all the pulses, while in the second case she applies \irud only to a fraction $1-(p_{n} K_{n})$ of the pulses, and \pns to the rest. 

In this case where DSP is applied the total keyrate can be expressed directly by equation \eqref{Ktot-compact}: 
$K = \sum_{n} \left[Y_{n} - \Pdelta \right]_{+} P_{n|\mu} K^{\text{marg}}_{n, m}$.

\begin{figure}
	\includegraphics[width=\columnwidth]{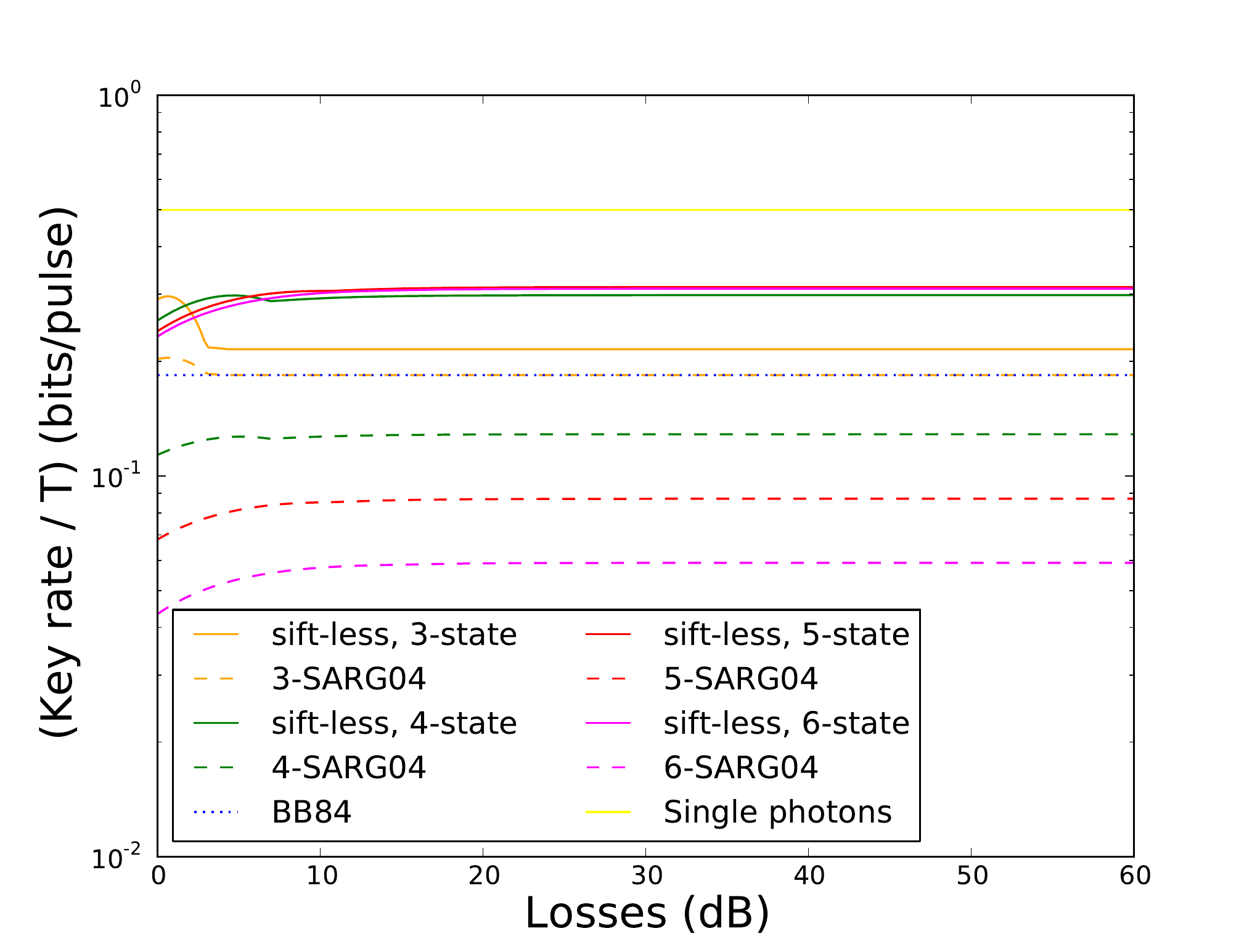}
	\caption{\label{FigDecoyKeyRates}
	(color online) Keyrates of the $m$-states protocols compared to $m$-state SARG04 and BB84
	with WCPs for $\mu=1$ and $m\in\llbracket3,6\rrbracket$, in the case the decoy-states protocol is applied. Since the behaviour is almost linear in T, we have plotted $K/T$, obtaining almost horizontal lines. }
\end{figure}

\begin{figure}
	\includegraphics[width=\columnwidth]{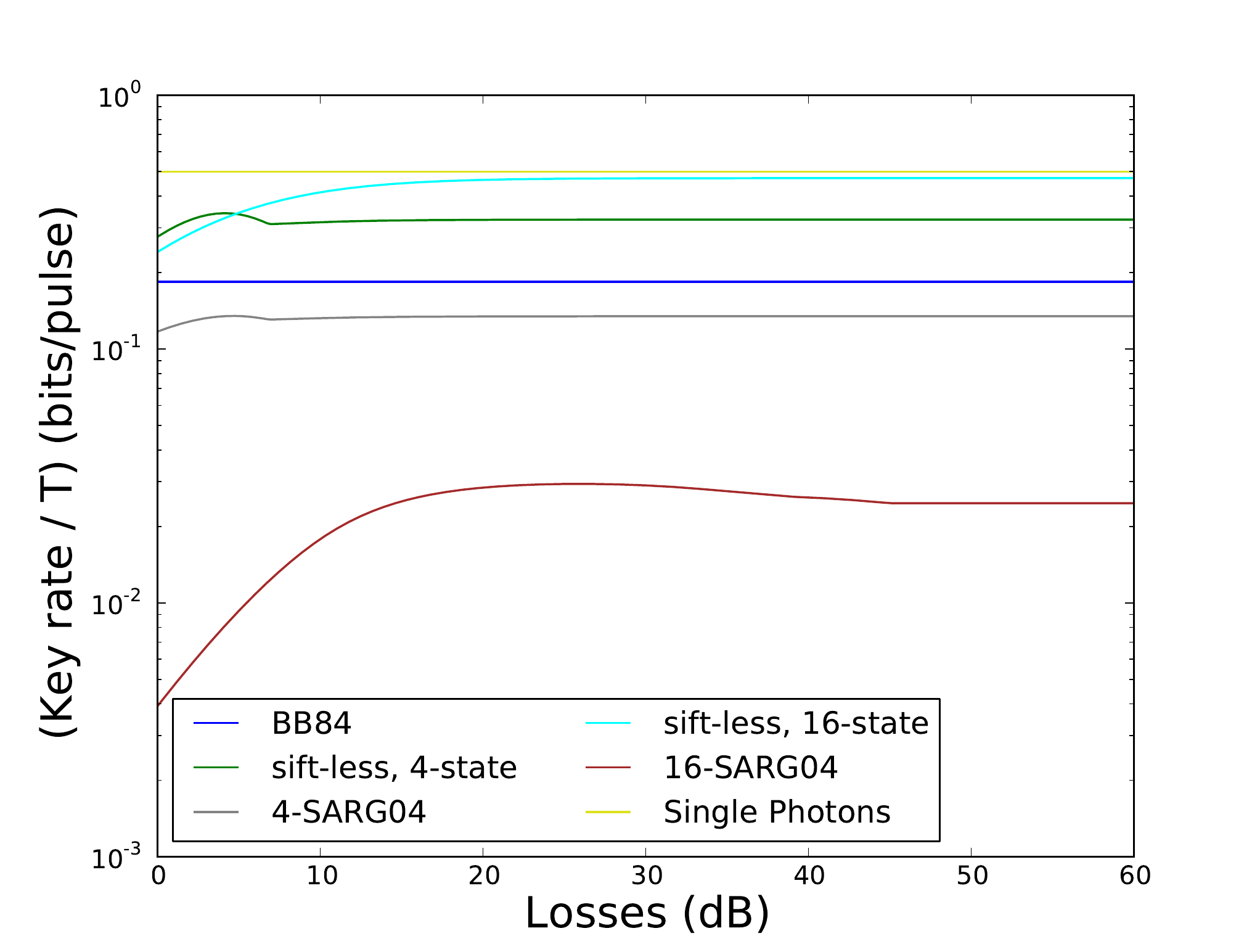}
	\caption{\label{FigDecoyOptKeyRates}
	(color online) Keyrates with optimized $\mu$, for the sifting-less protocol for $m=4$ and $m=16$, and for BB84, in the case DSP is applied. Again, to qualitatively prove that the behaviour of the keyrate is almost linear in T, we have  plotted the ratio $K/T$ instead of just the keyrate, obtaining almost horizontal lines.}
\end{figure}

  As shown below, these keyrates scale linearly with $T$. We have therefore plotted  $K(T)/T$ in figures \ref{FigDecoyKeyRates} and \ref{FigDecoyOptKeyRates} : the roughly horizontal shapes confirm  the  approximative proportionality to the transmission of the channel, either for BB84, for SARG and for the sifting-less protocol presented in this paper.
In figure \ref{FigDecoyKeyRates}  we show the keyrates for a fixed value of $\mu=1$ and in figure \ref{FigDecoyOptKeyRates} we have the keyrates optimized with respect to $\mu$, for each value of $T$.

\subsection{Comparisons between sifting-less protocol and BB84}
\label{comparisons-section}

In the case without DSP, the sifting-less and BB84 protocols scale differently with the transmission: 
for the sifting-less protocol we have seen how the optimized keyrate scales as $K \propto T^{1+ \frac{1}{m-2}}$ (see \eqref{k-opt})
where $T$ is the transmission;  the correspondent keyrate behaviour for BB84 is $K \propto T^{2}$.
This qualitative difference translates into important quantitative differences for low transmission rates.

In the case where DSP is applied to the BB84 protocol,
only single photons contribute to the keyrate, which is therefore
\begin{equation}
  K^{\text{BB84}}=\mu e^{-\mu}\frac{T^1}{1!}\frac{\log 2}2.
\end{equation}
A straightforward optimization allows to find 
$\mu_{\text{opt}}^{\text{BB84}} = 1$, and hence the keyrate is 
\begin{equation}
\frac{K_{\text{opt}}^{\text{BB84}}}{T}= \frac{\log 2}{2e}  = 0.1839\ \text{bits}.
\end{equation}
%
%
%

For the sifting-less protocol, if in \eqref{Ktot-compact} we explicit $K^{\text{marg}}_{n, m}$ as in \eqref{marg-keyrate} we obtain:
\begin{equation}\label{siftless-mufix}
\begin{split}
K^{\text{\textsc{DSP}}} &= \sum_{n=1}^{\infty} P_{n|\mu}Y_{n} K_{n}\\
&=\sum_{n=1}^{\infty} P_{n|\mu} [1-(1-T)^{n}] K_{n}.
\end{split}
\end{equation}

 Values of $n \geq m-1$ imply that  $\mathcal{P}(\Delta|n,m)\neq0$, then values of  $Y_{n}\leq \mathcal{P}(\Delta|n,m)$ are such that the contribution to the keyrate vanishes:   $K_{n}=0$. 
 Moreover, small values of $T$ imply small values of $Y_{n}$, and as figure \ref{Ktilde-n} shows, all the terms $n\geq m-1$ are null when $T$ is small enough. Therefore we can replace the infinite sum in \eqref{siftless-mufix} with a finite sum:
\begin{equation} \label{KDSP}
K^{\text{\textsc{DSP}}} =\sum_{n}^{m-2} P_{n|\mu} [1-(1-T)^{n}] K_{n}.
\end{equation}
The hypothesis $T\ll 1$ justifies also the approximation $1-(1-T)^{n} \approx n T$:
\begin{equation} \label{K-approx}
K^{\text{\textsc{DSP}}} \approx T \sum_{n}^{m-2} P_{n|\mu} n K_{n}.
\end{equation}
Finally, since we are in the case $m=4$ we can explicitly write the sum as a sum of only two terms:
\begin{equation} \label{k-decoy-m4-simple}
\frac{K_{m=4}^{\text{\textsc{DSP}}}}{T} \approx   e^{-\mu} [ \mu K_{1} +  \mu^{2} K_{2}  ].
\end{equation}
 
If we substitute the value $\mu=1$, which gives the maximum keyrate for the BB84 protocol, and use the values $K_{1} = 0.5$ bit/pulse and $K_{2} = h(\tfrac14)-\tfrac{\log 2}2 = 0.3113$ bits/pulse, we have:
\begin{equation}
\begin{split}
\frac{K_{m=4}^{\text{\textsc{DSP}}} (\mu=1)}{T}  &= \frac1e [ K_{1} + K_{2}]\\
& = 0.2987\ \text{bits}
\end{split}
\end{equation}
This result shows that even for the value of $\mu$
 optimized  for the BB84 protocol, the sifting-less protocol performs better, with an improvement of the keyrate of 62.26\%.

 We can also 
optimize the keyrate for the sifting-less protocol with respect to $\mu$, and compare it with the analog optimal value for BB84.
Computing the first derivative of \eqref{k-decoy-m4-simple} with respect to $\mu$ it's easy to show that this optimal value of the average photon number $\mu$ is $\mu_{\text{SLprot}}^{\text{opt}} = 1.4794$. Substituing this value in \eqref{k-decoy-m4-simple} we have the optimal value of the keyrate for the sifting-less protocol in the DSP case:
\begin{equation}
\frac{K_{m=4}^{\text{\textsc{DSP}}} (\mu^{opt})}{T} = 0.3237\ \text{ bits}.
\end{equation}
This optimal value of $\frac{K}{T}$ or the sifting-less protocol represents an improvement of  $75.96 \%$ over the analog optimal value for BB84 protocol.

From this starting observation, we can  look into two directions: 
we can  either consider what happens if $m$ is increased, or we can relax the $T\ll 1$ hypothesis.

\begin{figure}
	\includegraphics[width=\columnwidth]{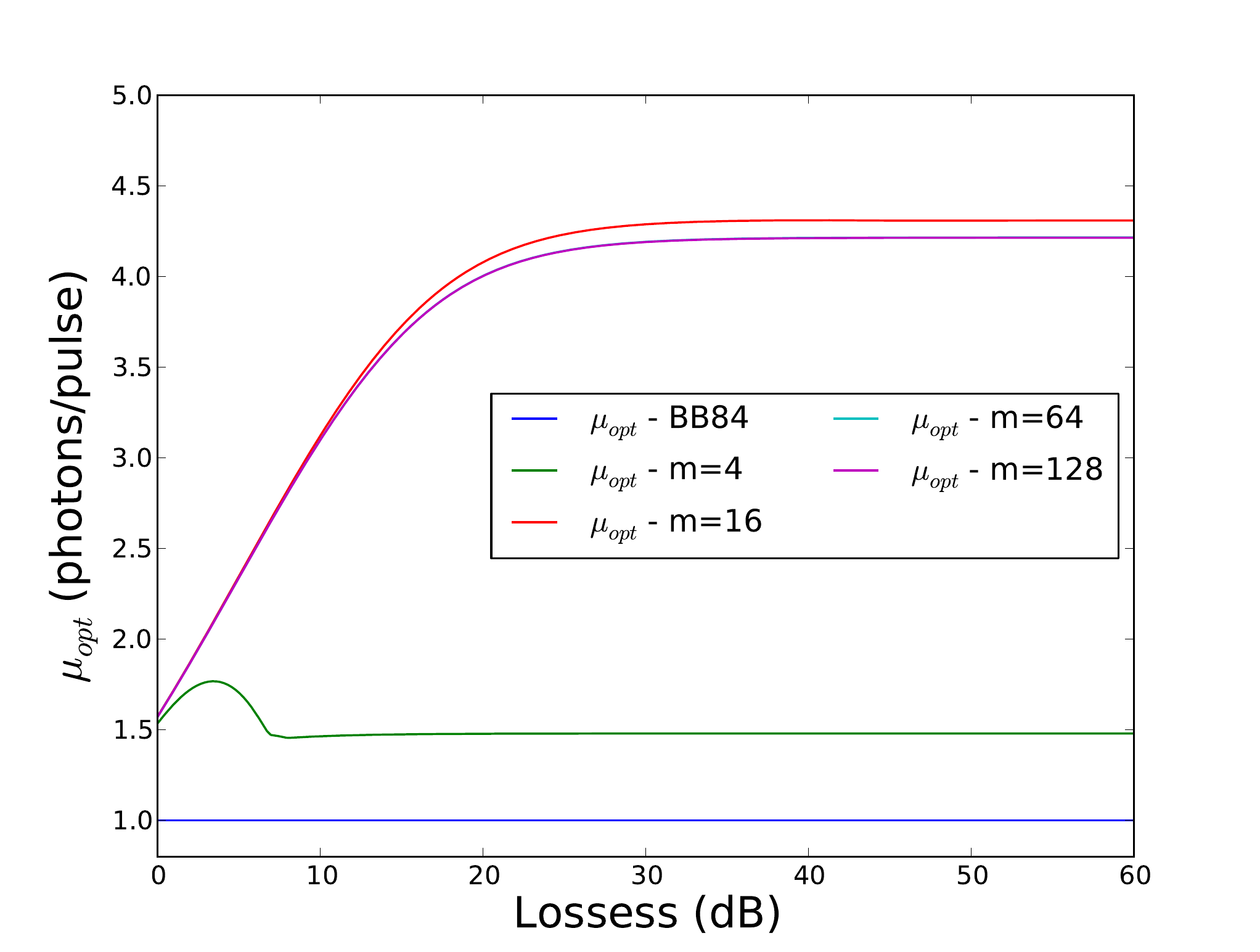}
	\caption{\label{figmuopt} (color online) Average photon number  $\mu_{\text{opt}}$  which optimizes the keyrate, as function of the attenuation of the channel, in the case DSP is applied. Different curves show $\mu_{\text{opt}}(T)$ for different values of $m$. For $m=4$, it is $\mu_{opt}\simeq 1.5$, in accordance with the qualitative analysis described in the text. The curves for $m=64$ and $m=128$ are almost superimposed, suggesting an asymptotic behavior. The optimal average photon number for BB84, in the  DSP case $\mu^{\text{BB84}}_{\text{opt}}(T)=\text{const}=1$ is shown for comparison.
	}
\end{figure}

For high values of $m$ (and still in the hypothesis of $T\ll1$) the protocol becomes closer to a continuous variables protocol.  Figures \ref{FigDecoyKeyRates} and \ref{FigDecoyOptKeyRates} show how, as $m$ increases, the sifting-less protocol keyrate  approaches  the keyrate for the ideal ``BB84 with perfect single photons'' protocol.

Figure \ref{figmuopt} shows a plot of numerically calculated values for $\mu_{\text{opt}}$ as function of the channel attenuation.

 In the right hand side of this figure (high losses, \ie  $T \ll 1$) we can observe that the numerically computed optimal average photon number stabilizes on a value of  $\mu_{\text{opt}}\simeq4.21$ as $m$ increases.

\begin{figure}
\centering
\includegraphics[width=\columnwidth]{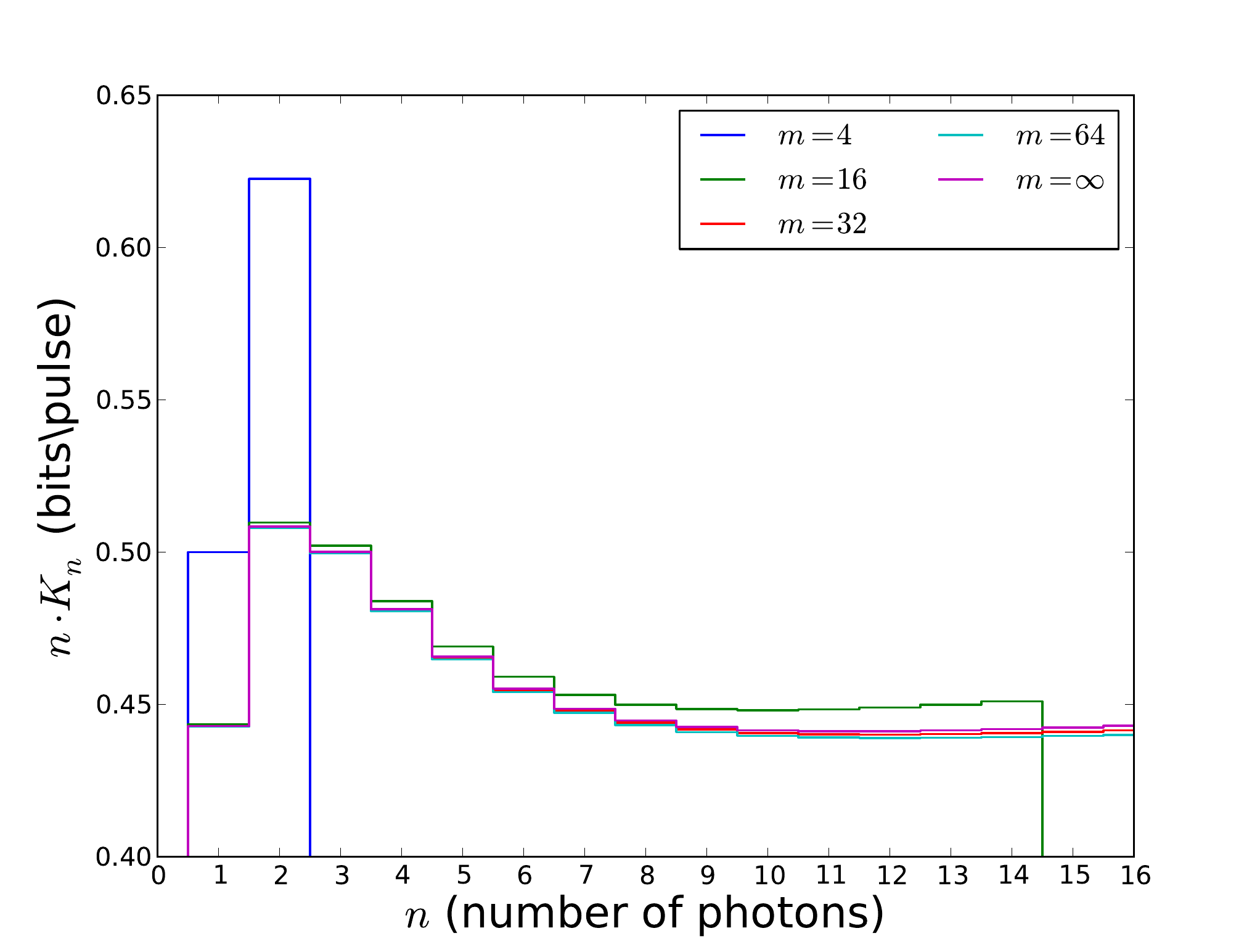}
\caption{\label{nKnvsn} (color online) Plot of the product $n \cdot K_{n}$ as function of $n$, for several values of $m$,
 and low values of $T$ (high attenuation), for the sifting-less protocol.  The plots are set to zero if the condition $n\geq m-1$ is met (see discussion above equation \eqref{KDSP}). The values of $K_{n}$ are given by formula \eqref{Kn}. Already for $m=64$ the behavior of $n\cdot K_{n}$ is roughly constant up to high values of $n$.  On the left we can observe the peak for low values of $n$. }
\end{figure}

Moreover, if we consider  increasing values of $m$ and small values of $T$, 
pulses with  increasing number of photons will have non zero contribution to the keyrate.
To have a qualitative intuition of the origin of this effect, we can use the  result (found with numerical computation), shown in figure \ref{nKnvsn} : the product $n \cdot K_{n}$ is roughly constant as function of $n$, in the limit case $m\rightarrow \infty$, and in the case of $m$ finite but big, it stays constant as long as $n\le m-2$.

Using this result, we can extend the qualitative analysis done for  $m=4$. Looking at the expression \eqref{K-approx} of the keyrate, we will  have $m-2$ non-zero terms, with the $P_{n|\mu}$ coefficients multiplying the roughly constant $n \cdot K_{n}$ product.  On the other hand, in figure \ref{nKnvsn} we have a pronounced peak centered between 2 and 3 photons, for small values of $m$ ($m=4$), and this explains the lower value of $\mu_{opt}$ for $m=4$ in figure \ref{figmuopt}. For higher values of $m$, the shape of the plots in figure \ref{nKnvsn} are roughly similar, with a less pronounced peak, shifted toward slightly higher values of $n$. This explains the fact that in figure \ref{figmuopt}, $\mu_{opt}$ stabilizes (for high losses) on the value of $4.21$ as $m \to \infty$.

Once again, this behaviour shows the advantage of the sifting-less protocol: pulses with high photon number have a non-negligible contribution to the keyrate, \ie   multiple-photon-pulses are more robust to PNS attacks.

If we now consider higher values of $T$, we have a similar effect on the keyrate: even for small values of $m$, as $T$ increases we have  increasing values of $n$ with non-zero contribution to the keyrate, 
as can be seen in  figures \ref{Ktilde-n} and \ref{rates-figure}. 
On the other hand figure \ref{rates-figure} shows how for higher values of $n$ the value of $K_{n}$ decreases, meaning that even if pulses with high number of photons have a non-zero contribution, this contribution becomes smaller with $n$. 

Again, those behaviors  show the higher efficiency of the sifting-less protocol, and its robustness against PNS attacks.

\section{Conclusions}
\label{secConclusion}
The sifting-less protocols described here are as efficient as BB84 and more
robust against PNS-attack,
and are compatible with DSP protocols. 
This robustness 
lies in the preservation of non-orthogonality of the sent-states
by the lack of sifting.

Furthermore, this also allows to 
extract a reasonable key for high $m$, while benefiting from the
robustness brought by the increased overlap of the sent states, 
contrarily to the $m$-state SARG04 variant, which while robust, has a sifting 
factor $\propto m^{-3}$ \cite{SARG2}.

The most robust variant limit of this protocol is the limit of 
continuous phase modulation $m\rightarrow\infty$, which actually prevents
the IRUD attack. It is straightforward to show that replacing the $m$-state POVM
used in the above description with the simpler 4-State POVM used in standard BB84
does not change the
keyrate in this limit. 
This really shows that the robustness comes
from the non-orthogonality of the sent-states, which is preserved by the
lack of sifting.

Before using this protocol, we still need to investigate its security 
in presence of a non-zero QBER. 
For perfect single photons and a QBER $\epsilon$, 
one can bound Eve's information by writing the state
shared by Alice, Bob and Eve under the form \cite{SBPCDLP09}
$\ket{\Psi_{ABE}}
  =\sqrt{\lambda_1}\ket{\Phi^+}\ket{E_1}
  +\sqrt{\lambda_2}\ket{\Phi^-}\ket{E_2}
  +\sqrt{\lambda_3}\ket{\Phi^+}\ket{E_3}
  +\sqrt{\lambda_4}\ket{\Phi^-}\ket{E_4}$,
and optimizing Eve's Holevo information $S(Y{:}E)$. 
One then straightforwardly find $S(Y{:}E|n=1,\epsilon)=h(\epsilon)$.
For $m=4$, we have 
$S(X{:}Y)=\tfrac12(\log2-h(\epsilon))$, which gives a net keyrate 
$K=\frac12(\log2 - 3 h(\epsilon))$, cancelling for a QBER  $\epsilon=6.14\%$.
The expression is less elegant for other values of $m$, but the critical value of
$\epsilon$ does not change much, varying between 6.89\% for $m=3$ and
5.93\% for $m\rightarrow\infty$. 
Of course, for a practical application of these protocols, the combination of 
QBER and PNS attacks still needs to be investigated, 
as well as finite-size effects 
\cite{CaiScarani09},
and the robustness of the system to imperfections in the data processing 
and in the experimental set-up \cite{VakhitovMakarovHjelme01, QiFungLoMa07, LydersenWiechersWittmann+10}. 
Another direction worth investigating would be an unbalanced version of our 
protocol, similar to BB84 with biased basis choice \cite{LoChauArdehali98-05, Ardehali-98}, allowing 
to double the keyrate to $\sim1$ bit instead of $\sim.5$ in the low-loss 
regime.

\appendix

\begin{acknowledgments}

Frédéric Grosshans thanks Valerio Scarani, for bringing the problem of the optimal
sifting of the states used in BB84 and SARG04 to his attention during a visit
at the Centre for Quantum Technologies at the National University of Singapore.
This research has been funded by the European Union under the EQUIND
(project {IST-034368}) and NEDQIT (ERANET Nano-Sci) projects,
by the French Agence Nationale de la Recherche PROSPIQ and FReQueNCy projects
(projects ANR-06-NANO-041 and ANR-09-BLAN-0410) and by the 
NSERC FReQuenNCy project (R0018554). 

\end{acknowledgments}

\bibliography{biblio}

\end{document}